\documentclass[12pt,onecolumn,showpacs,amssymb,aps,nofootinbib,floatfix]{revtex4-1}

\usepackage{epsfig}
\usepackage{color}
\usepackage{latexsym}
\newcommand{\req}[1]{Eq.\,(\ref{#1})}

\newcommand{\beq}{\begin{equation}}
\newcommand{\eeq}{\end{equation}}

\newcommand{\eqcomma}{,\phantom{AA}}
\newcommand{\order}[1]{ \mathcal{O} \left( #1 \right) }
\newcommand{\ave}[1]{\left\langle #1 \right\rangle}

\begin{document} 
\hbadness=10000

\title{The Unruh effect and oscillating neutrinos}

\author{Dharam Vir Ahluwalia$^a$, Lance Labun$^b$, Giorgio Torrieri$^c$}
\affiliation{
$^a$Department of Physics, Indian Institute of Technology,
Kalyanpur, Kanpur, Uttar Pradesh 208016, India\\
and Department of Physics and Astronomy, University of Canterbury, Christchurch 8140, New Zealand
\\
{\rm email:dharam.vir.ahluwalia.1952@gmail.com}
\\
$^b$ Department of Physics, University of Texas,  Austin, TX 78712, USA\\
{\rm email:lance@phys.ntu.edu.tw}\\
$^c$ IFGW, Universidade Estadual de Campinas, Campinas, S$\tilde{a}$o Paulo, Brazil\\
{\rm email:lunogled@gmail.com}\\
Essay written for the Gravity Research Foundation 2015 Awards for Essays on Gravitation}

\date{\today}

\begin{abstract}
We point out that neutrino oscillations imply an ambiguity in the definition of the vacuum and the coupling to gravity, 
with experimentally observable consequences due to the Unruh effect.   In an accelerating frame, the detector should see a bath of mass Eigenstates neutrinos.   In inertial processes, neutrinos are produced and absorbed as charge Eigenstates.  The two cannot be reconciled by a spacetime coordinate transformation.   This makes manifestations of the Unruh effect in neutrino physics a promising probe of both neutrinos and fundamental quantum field theory.
In this respect, we suggest $p\rightarrow n +\ell^+ + {\nu_\ell}$ ($\ell^+ = e^+, \mu^+, \tau^+$) transitions in strong electromagnetic fields as a promising avenue of investigation.
In this essay we discuss this process both in the inertial and comoving frame, we briefly describe the experimental realization and its challenges, and close by speculating on possible results of such an experiment in different scenarios of fundamental neutrino physics
\end{abstract}


\maketitle
Quantum field theory has justifiably been at the forefront of fundamental physics for the last few decades.    The physical world in this picture is represented by a Lorentz invariant manifold (spacetime) on which point-like quantum particles, described by irreducible representations of the Lorentz group, propagate and interact locally \cite{weinberg}. 
This formalism has described properties and interactions of fundamental particles and forces to excellent precision, with two notable exceptions.

The first exception is gravity, which considered from the point of view of QFT still presents both formidable technical difficulties \cite{weinberg3,dharam,gteq} and fundamental conceptual paradoxes.   This investigation so far has mostly been purely theoretical, since general dimensional analysis arguments \cite{dharam} indicate quantum-gravitational effects are $\order{k/10^{19}\,\mathrm{GeV}}$ where $k$ is the characteristic momentum scale.   This is generally a formidable barrier.

The second exception is neutrino oscillations.   While a phenomenological description of them has been known for decades \cite{pontecorvo} and a definition of a propagator was developed \cite{blasone,naumov}, this model has not been incorporated into a rigorous field theory, since the vacuum of the neutrino field is ambiguous.    This is in contrast to the superficially similar meson mixing, successfully incorporated into the standard model \cite{weinberg}. A fundamental difference between the two cases is that meson mixing is due to the non-commutativity of the strong and electroweak currents, and all mixing meson states are mass-degenerate.
In contrast, neutrinos apparently have no superselection rule forbidding mass mixings.     In practice, since the right handed neutrino has no non-gravitational interactions, so far neutrinos could be created only in their charge Eigenstates, and subsequently measured after propagation as other charge Eigenstates, with mass Eigenstates not being directly producible (the experimental program proposed here could change this).

To see the implications for gravitational physics \cite{olddharam,blasmag,ourpaper}, consider neutrinos interacting both with weak gravitational fields and with electroweak  gauge fields.   Metric perturbations couple to the energy-momentum tensor $\hat{T}_{\mu \nu}$, which depends on the mass.   Weak interactions, on the other hand, couple to electroweak current operator $\hat{J}^\mu$, which defines the charge eigenstates.
The Pontecorvo mechanism makes  local conservation of these currents impossible to satisfy simultaneously since the energy-momentum tensor $\hat{T}_{\mu \nu}$ and the weak currents $\hat{J}_\mu$ do not commute even in the free field probe limit (neglecting any weak interaction terms $\sim \hat{J}\times \hat{J}$ contributions to the Lagrangian).  Written in terms of mass basis fields $\psi_j$, the electroweak and 4-momentum currents of a free neutrino-only field are  
\begin{equation}
\label{jdef}
\hat{J}_{Ll}^\mu \simeq \sum_{j=1,..,3} U_{lj} \left( \overline{\psi_L}_{j} \gamma^\mu \psi_{Lj} \right) \eqcomma \hat{T}_{\mu \nu}= \sum_{i=1...3}
  \left [\overline{\psi_{i}} \gamma_\mu \partial_{\nu} \psi_{i} - g_{\mu \nu} \left( \overline{\psi}_i (\gamma_\alpha \partial^\alpha
-m_i) \psi_i \right) \right] + \order{h,\psi^4}
\end{equation}
$U_{lj}$ is a unitary $3 \times 3$ matrix connecting weak charge and mass Eigenstates usually called the Pontecorvo-Maki-Nakagawa-Sakata matrix \cite{valle}.  
 It is clear that an eigenstate of $\hat{T}_{\mu \nu}$ is not an eigenstate of $\hat{J}_\mu$  and vice versa. 

For all other fields except neutrinos, in the free field limit, the only measurement that does not commute with $\hat{T}_{\mu \nu}$ is the field strength at a given position, $\hat{\psi}(x)$.  Provided the $\ave{ g_{\mu \nu} \Delta \hat{T}^{\mu \nu}}$ uncertainity triggered by a position measurement is $\ll l_p^{-2} R$, where $R$ is the scalar curvature of spacetime and $l_p$ the Planck length, quantum uncertainity will not affect the definition of locality.
  Thus, in the weak-field small-curvature limit general relativistic covariance has a well-defined semiclassical limit.   
  
Neutrinos, due to \req{jdef}, provide another way to test for the impact of quantum interactions on gravity.  Weak charged current interactions couple to the weak charge of the neutrino by involving a charged lepton and in this way ``instantly'' modify the energy momentum tensor.  Unlike with the position measurement case, this modification will happen when the neutrino interacts  at a momentum scale many orders of magnitude smaller than $l_p^{-1}$.  Consequently, the change of $\hat{T}_{\mu \nu}$ due to weak interactions
\begin{equation}
\ave{T_{\mu \nu} (\psi_{Li},\psi_{Ri})} \rightarrow\sum_j U_{ij} \ave{T_{\mu \nu} \left(\psi_{Lj},\psi_{Ri} \right)}
\end{equation}
can be significant also in the weak curvature limit.
For a neutrino-self gravitating system will have instantaneous (and in the semiclassical limit) non-causal effects on geometry
(this is similar to frustrated equilibrium \cite{frust} systems in condensed matter physics) 

\begin{figure}[h]
\epsfig{width=0.84\textwidth,figure=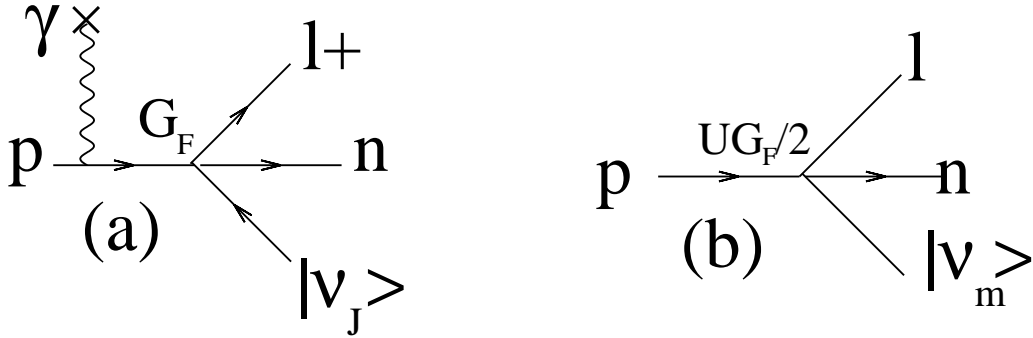}
\caption{\label{diagram} 
What an accelerated $p \rightarrow n l \overline{\nu}$ process looks like in a comoving frame (panel (b))  and an inertial frame
(panel (a)).  In the semiclassical limit the $\gamma$ line means a classical external electromagnetic potential, treated nonperturbatively, rather than perturbation theory (the $\nu$ and $l$ lines do not have arrows as they can be either absorbed from the Unruh vacuum or emitted, corresponding to the three different comoving processes in \cite{vanzella1})
}
\end{figure}

The gravitational interaction of neutrino matter is not a feasible measurement in the foreseeable future, though it might have consequences in astrophysical settings.    
In this work, we examine the counter-part in {\em accelerating} rather than gravitational frames via the Unruh effect \cite{unruh,matsasrev,unruhweiss}.  There, the curvature is zero and the metric is generated from the flat space metric by an infinitesimal Rindler deformation.
The creation and annihilation operators of {\em mass} Eigenstates are different in this coordinate system \cite{unruh}: the accelerating probe sees a thermal distribution of {\em mass} states, whose occupation numbers at zero chemical potential depend on energy-momentum only.   The inertial equivalent, an analogue of Bremstrahlung in electroweak theory \cite{vanzella1} (fig. \ref{diagram} left panel), requires that neutrinos emitted and absorbed by the proton be Eigenstates of the weak charge $\hat{J}$, i.e. left-handed flavor Eigenstates $\nu_{e,\mu,\tau L}$.
 As \req{jdef} shows both mass Eigenstates and charge Eigenstates are defined by Lorentz-invariant local operators, they cannot be related by a coordinate change.
Since the Unruh effect is a universal feature of quantum field theories \cite{biso,sewell},
this yields an ambiguity, directly related to the ambiguity of the neutrino vacuum, that could be investigated in the laboratory.

To illustrate the problem better, it needs to be remembered \cite{vanzella1,vanzella2,suzuki,higuchi1,higuchi2,sudarsky} that the Unruh effect could be interpreted as the quantum field theory analog of ``coriolis forces'':   an artifact of viewing a process defined within a theory invariant under special relativity in a noninertial frame.  For weak interactions, this was shown explicitly by \cite{vanzella1,vanzella2,suzuki} through  the conversion of accelerated protons into neutrons, $p \rightarrow n e^+ \nu$, illustrated in Fig. \ref{diagram}.  The inertial observer sees a proton, accelerated for example by a classical electromagnetic field, decay into a neutron, a positron and a neutrino.   Quantum mechanically, the electromagnetic field can be represented as a background field imparting virtuality to the proton, making the Bremsstrahlng-like process kinematically possible.
In the comoving frame, an observer sees a thermal bath of the light particles involved in the decay:  neutrinos, electrons and positrons.   The conversion rate is given by the cross section of the proton colliding with particles in the thermal bath.

\cite{vanzella1,vanzella2,suzuki} have shown these processes to be equivalent at tree level  in the low acceleration limit treating the masses as fundamental, the interaction as a Fermi contact interaction and neglecting neutrino oscillations.  We now see that lifting the last simplification by using the physical neutrino spectrum introduces a fundamental ambiguity in the calculation, which arises directly from the ambiguity inherent in a quantum field theory of oscillating neutrinos.   

\begin{figure}[h]
\epsfig{width=0.94\textwidth,figure=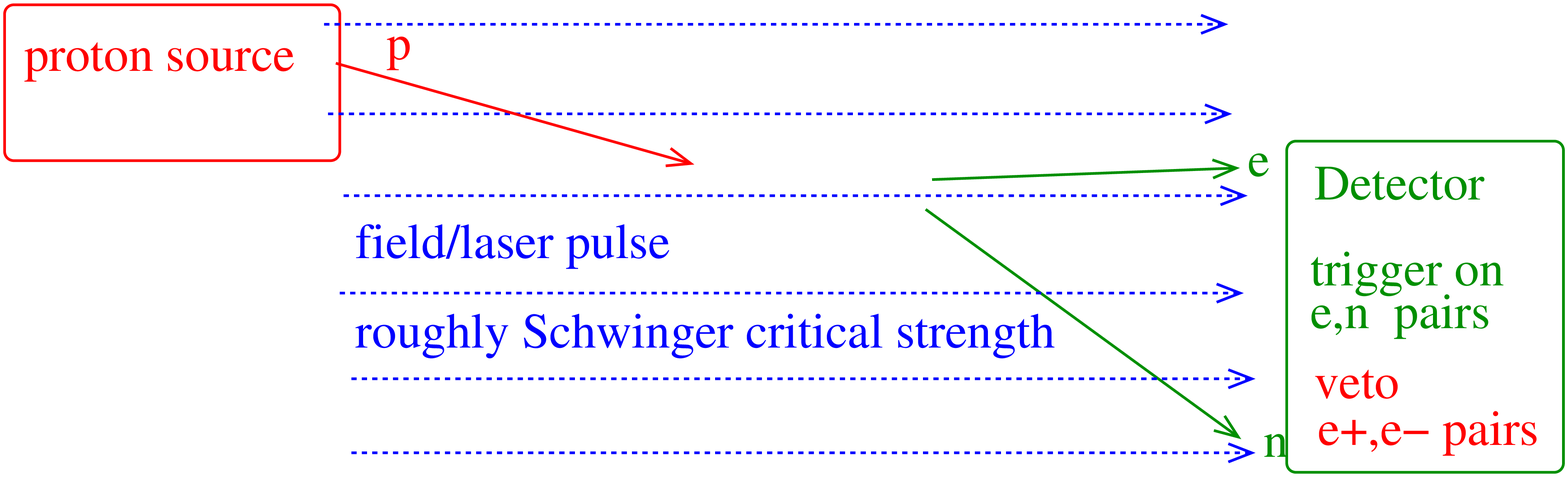}
\caption{\label{setup}
A sketch of the experimental setup required for the measurement of the $p\rightarrow n \nu$ conversion. The technical specifications will be described in \cite{ourpaper}
}
\end{figure}

This calls for an experimental investigation of $p \rightarrow n e^+$ conversions. The challenges for such an investigation are considerable but may be surmountable.  For semiclassical acceleration to be well-defined, we need a strong approximately constant electromagnetic field.   An order of magnitude estimate shows \cite{ourpaper} that
experimental setups capable of investigating the Schwinger effect in strong electromagnetic fields \cite{dunne,eli} will also be capable of investigating the phenomenon discussed here, by adding 
proton beam injection and neutron detectors with sufficient triggering precision to look for neutron-positron coincidences and veto Schwinger-generated electron-positron coincidences (Fig. \ref{setup}).

  A major advantage is that we do {\em not} have to detect any neutrinos to infer a conversion, which comes as a time-coincident $n e^+$ pair without a corresponding $e^-$.
 The observable we are interested in is the differential production rate and its dependence on the proton acceleration and initial momentum.   The experimental observable can be compared to the prediction of both the inertial and comoving calculation in \cite{vanzella1,vanzella2}.  

\begin{figure}[h]
\epsfig{width=0.44\textwidth,figure=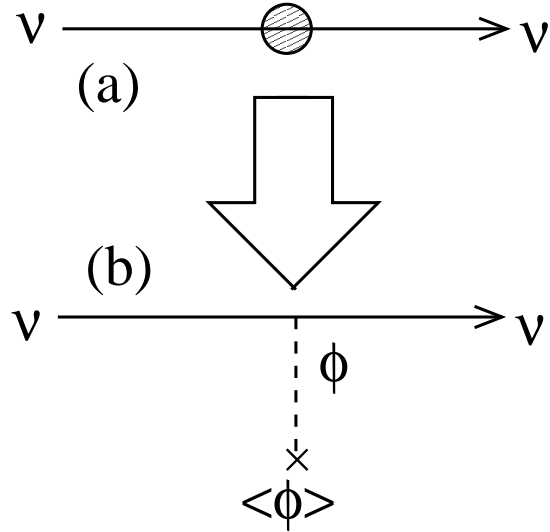}
\caption{\label{higgsdiagram} 
UV picture of mass generation via the Higgs mechanism.   For an Unruh effect
analysis of Bremsstrahlung at accelerations comparable to the Higgs condensate density in natural units, diagram (b) needs to be transformed into the Rindler frame.}
\end{figure}

While we refer the reader to \cite{ourpaper} for technical details, it is clear that the mixing and possible presence of right-handed neutrinos ensure the inertial and comoving predictions examined in \cite{vanzella1,vanzella2,suzuki} will differ by an approximately momentum-independent normalization factor dominated by $U_{1e}/2$.   This implies {\em at least one} of these calculations will disagree with experimental measurement.
To find out which one and connect this issue to the origin of the neutrino mass, we must understand that the inertial and the comoving frame calculation assume point-like particles with ``fundamental'' masses, represented by a diagram such as
Fig. \ref{higgsdiagram} panel (a).   However, in the standard model masses are IR operators  (Yukawa fermion couplings to the Higgs condensate) vanishing in the ultraviolet \cite{weinberg}, and it is possible that the discrepancy in Rindler-frame and inertial-frame results is resolved by understanding origin of neutrino mass.
Seeing that the inconsistency of outcomes arises when assuming point-like "fundamental" masses rather than effective terms, one straight-forward explanation is that this approximation breaks down in different ways in the two frames of reference, meaning one of the two calculations is inappropriate.

The most common hypothesis is that the neutrino mass is not fundamental, but produced either  via interaction with a Higgs or Higgs-like condensate, $\Delta L_{dirac} \sim g \ave{\phi} \nu \overline{\nu}$, or via a higher dimensional operator (e.g., mixing with a hypothetically massive right handed Majorana neutrino, 
$\Delta L_{majorana} \sim (M^2/\Lambda_{uv} ) \nu \overline{\nu}$
where $M$ is a scale similar to other standard model masses and $\Lambda_{uv}$ is some high energy scale, explaining the smallness of the mass).

In both cases, while the scale of the physics responsible for the mass is far higher than any conceivable experimentally explorable acceleration, {\em the mass itself} is extremely low, comparable in natural units to accelerations obtainable in the laboratory.  Since by dimensional analysis this is the relevant operator required for the ``point-like mass'' EFT to break down, the experiment proposed here can be seen as a window on the origin of neutrino masses.

In such scenarios, the approximations made in \cite{vanzella1,vanzella2} are more robust in the inertial than in the comoving calculation.    If the mass is due to a condensate, it will appear in the accelerating frame as a Rindler-transformed bath of zero-frequency scalars (Fig. \ref{higgsdiagram} (b)).    The formal identity between the thermal and the Unruh vacuum indicated in works such as \cite{sewell} and \cite{unruhweiss} would imply that the $\ave{\phi}$ condensate  and the coupling constant $g$ are renormalized to values  depending on the effective temperature, analogously, but at a much lower energy scale,  to the $\sigma$-model calculation of \cite{ohsaku,zhuk,castorina}.   The comoving calculation in \cite{vanzella1,vanzella2} does not exhibit such a condensate, and hence it is not a good approximation to the neutrino mass, as measured by the comoving observer, if a condensate is the origin of this mass.

However, the robustness of the inertial calculation is not guaranteed.
   An intriguing possibility is that the charge Eigenstate is due to infrared physics vanishing in the UV \cite{bob}.    In this case, the non-inertial Unruh calculation might be the correct one, while the Bremsstrahlung calculation would break down because the neutrino-quark coupling operator would be modified by IR effects.  While such an alternative is not often discussed in the literature, constructions of this type do exist \cite{bob}.  This scenario, signalled by the Bremsstrahlung calculation being incorrect by a factor of $2 U_{ij}^{-1}$, would be invaluable for studying neutrino oscillations:   unlike the  present neutrino oscillation measurements \cite{valle}, this measurement would be sensitive not to phase differences, but the {\em absolute values} of masses and mixing angles.   The possibility of this scenario is a good motivation for studying the feasibility of this experimental program further.

More intriguingly, the inconsistency might signal the breakdown of the basic formalism of quantum field theory in which {\em both} the inertial and comoving calculation break down.    At the moment this scenario is too poorly explored on the theoretical level to make even qualitative predictions, but perhaps tests of CPT invariance (perhaps with strongly accelerated antiprotons as well as protons) can extricate this alternative convincingly.
Probing the Unruh effect gives us the ability to prepare and probe mass as well as charge Eigenstates,  thereby doubling the number of experimentally accesible constraints we have at our disposal for fundamental symmetry tests.
 
In conclusion, the conversion of accelerated protons into neutrons is a promising laboratory to study the quantum field theoretical origin of neutrino masses and oscillations, as well as more generally the relationship between the spacetime manifold and the pointlike particles traveling on it.
We hope that the ingenuity of the QED and intense laser community will make such experiments feasible in our lifetimes.

\begin{thebibliography}{19}


\bibitem{weinberg} S. Weinberg,The quantum theory of fields, volume I,II


\bibitem{weinberg3}
  S.~Carlip,
  Rept.\ Prog.\ Phys.\  {\bf 64}, 885 (2001)
  [gr-qc/0108040].

\bibitem{dharam} 
  D.~V.~Ahluwalia,
  Mod.\ Phys.\ Lett.\ A {\bf 17}, 1135 (2002)
  [gr-qc/0205121].



\bibitem{gteq} 
  G.~Torrieri,
  arXiv:1501.00435 [gr-qc].


\bibitem{pontecorvo} 
  B.~Pontecorvo,
  Sov.\ Phys.\ JETP {\bf 26}, 984 (1968)
  [Zh.\ Eksp.\ Teor.\ Fiz.\  {\bf 53}, 1717 (1967)].

\bibitem{blasone} 
  M.~Blasone, F.~Dell'Anno, S.~De Siena and F.~Illuminati,
  Europhys.\ Lett.\  {\bf 106}, 30002 (2014)
  [arXiv:1401.7793 [quant-ph]].

\bibitem{naumov} 
  D.~V.~Naumov and V.~A.~Naumov,
  J.\ Phys.\ G {\bf 37}, 105014 (2010)
  [arXiv:1008.0306 [hep-ph]].


\bibitem{olddharam} 
  D.~V.~Ahluwalia,
  Mod.\ Phys.\ Lett.\ A {\bf 13}, 2249 (1998)
  [hep-ph/9807267].

\bibitem{blasmag} 
  M.~Blasone, J.~Magueijo and P.~Pires-Pacheco,
  Europhys.\ Lett.\  {\bf 70}, 600 (2005)
  [hep-ph/0307205].


\bibitem{ourpaper} D.Ahluwalia, L. Labun, G. Torrieri, The Unruh effect and neutrino oscillations, in progress

\bibitem{valle} 
  J.~W.~F.~Valle,
  J.\ Phys.\ Conf.\ Ser.\  {\bf 53}, 473 (2006)
  [hep-ph/0608101].

\bibitem{frust}
Alexei Tsvelik, ``Quantum Field Theory in Condensed Matter Physics'', 2007

\bibitem{unruh}
  W.~G.~Unruh,
  Phys.\ Rev.\  D {\bf 14}, 870 (1976).


\bibitem{matsasrev} 
  L.~C.~B.~Crispino, A.~Higuchi and G.~E.~A.~Matsas,
  Rev.\ Mod.\ Phys.\  {\bf 80}, 787 (2008)
  [arXiv:0710.5373 [gr-qc]].


\bibitem{unruhweiss} 
  W.~G.~Unruh and N.~Weiss,
  Phys.\ Rev.\ D {\bf 29}, 1656 (1984).


\bibitem{biso}
  J.~J.~Bisognano and E.~H.~Wichmann,
  J.\ Math.\ Phys.\  {\bf 17}, 303 (1976).

\bibitem{sewell}
  G.~L.~Sewell,
  Annals Phys.\  {\bf 141}, 201 (1982).



\bibitem{vanzella1} 
  D.~A.~T.~Vanzella and G.~E.~A.~Matsas,
  Phys.\ Rev.\ Lett.\  {\bf 87}, 151301 (2001)
  [gr-qc/0104030].

\bibitem{vanzella2} 
  D.~A.~T.~Vanzella and G.~E.~A.~Matsas,
  Phys.\ Rev.\ D {\bf 63}, 014010 (2001)
  [hep-ph/0002010].

\bibitem{suzuki} 
  H.~Suzuki and K.~Yamada,
  Phys.\ Rev.\ D {\bf 67}, 065002 (2003)
  [gr-qc/0211056].

\bibitem{higuchi1} 
  A.~Higuchi, G.~E.~A.~Matsas and D.~Sudarsky,
  Phys.\ Rev.\ D {\bf 45}, 3308 (1992).

\bibitem{higuchi2} 
  A.~Higuchi, G.~E.~A.~Matsas and D.~Sudarsky,
  Phys.\ Rev.\ D {\bf 46}, 3450 (1992).

\bibitem{sudarsky} 
  P.~Igor and S.~Daniel,
  arXiv:1306.6621 [quant-ph].


\bibitem{dunne} 
  G.~V.~Dunne,
  Eur.\ Phys.\ J.\ D {\bf 55}, 327 (2009)
  [arXiv:0812.3163 [hep-th]].


\bibitem{eli} The Extreme Light Infrastructure (ELI) project: http://www.extreme-light-infrastructure.eu/eli-home.php

\bibitem{ohsaku} 
  T.~Ohsaku,
  Phys.\ Lett.\ B {\bf 599}, 102 (2004)
  [hep-th/0407067].

\bibitem{zhuk} 
  D.~Ebert and V.~C.~Zhukovsky,
  Phys.\ Lett.\ B {\bf 645}, 267 (2007)
  [hep-th/0612009].

\bibitem{castorina} 
  P.~Castorina and M.~Finocchiaro,
  J.\ Mod.\ Phys.\  {\bf 3}, 1703 (2012)
  [arXiv:1207.3677 [hep-th]].

\bibitem{bob}
  B.~McElrath,
  arXiv:0909.3090 [hep-ph].\\
While this particular work was withdrawn, the notion that a Majorana mass can be a spin-mixing
operator induced by a finite chemical potential is feasible 


\end{thebibliography}
\end{document}